\documentclass[12pt]{article}
\usepackage{amsmath,amssymb}
\usepackage{graphicx}
\newcommand{\eqdef}{\stackrel{\text{def}}{=}}
\newcommand{\n}{\nonumber}
\newcommand{\bm}{\boldsymbol}

\newcommand{\ignore}[1]{}
\numberwithin{equation}{section}
\newcommand{\Romannumeral}[1]{\uppercase\expandafter{\romannumeral#1}}
\newcommand{\I}{\text{\Romannumeral{1}}}
\newcommand{\II}{\text{\Romannumeral{2}}}

\newcommand{\ppm}{\hspace{0.3em}\raisebox{0.5ex}{$>$}\hspace{-0.75em}\raisebox{-.6ex}{$<$}\hspace{0.3em}}

\allowdisplaybreaks[4]

\begin{document}

\baselineskip=20pt

\newfont{\elevenmib}{cmmib10 scaled\magstep1}
\newcommand{\Title}[1]{{\baselineskip=26pt
   \begin{center} \Large \bf #1 \\ \ \\ \end{center}}}
\newcommand{\Author}{\begin{center}
   \large \bf C.-L. Ho${}^1$, J.-C. Lee${}^2$ and R. Sasaki${}^{3,4}$ \end{center}}
\newcommand{\Address}{\begin{center}
     $^1$ Department of Physics, Tamkang University,
     Tamsui 251, Taiwan (R.O.C.)\\
     $^2$  Department of Electrophysics, National Chiao-Tung University, Hsinchu, Taiwan (R.O.C.) \\
     $^3$ Center for Theoretical Sciences,
    National Taiwan University, Taipei, Taiwan (R.O.C.)\\
    $^4$ Department of Physics, Shinshu University,
     Matsumoto 390-8621, Japan
   \end{center}}

\thispagestyle{empty}

\Title{Scattering Amplitudes for Multi-indexed Extensions of Solvable Potentials}

\Author

\Address
\vspace{1cm}

\begin{abstract}

New solvable one-dimensional quantum mechanical scattering problems are presented.
They are obtained from known solvable potentials by multiple Darboux transformations in terms of virtual 
and pseudo virtual wavefunctions.
The same method applied to confining potentials, {\em e.g.\/} P\"oschl-Teller and the radial oscillator potentials,
has generated the {\em multi-indexed Jacobi and Laguerre polynomials\/}.
Simple multi-indexed formulas  are derived for the transmission and reflection amplitudes of several
solvable potentials.

\end{abstract}

\section{Introduction}
We address the problem of rational extensions of solvable one-dimensional quantum mechanical 
scattering problems.
Study of exactly solvable potentials in one dimensional quantum mechanics \cite{infhul}--\cite{susyqm}
has been   rapidly developing in recent years.
It culminated in the discovery of multi-indexed Jacobi and Laguerre polynomials \cite{os25,gomez3}
together with the exceptional orthogonal polynomials \cite{gomez}--\cite{ho} as the main part of
the eigenfunctions for the rationally extended P\"oschl-Teller and radial oscillator potentials.
The main focus of these papers \cite{os25}--\cite{ho} has been iso-spectral deformations of the 
(semi) confining potentials and the bound state eigenfunctions.
These rational extensions (deformations) are achieved by multiple Darboux transformations 
\cite{darb,crum,adler} in terms of {\em polynomial type  seed solutions\/}, which are called
the {\em virtual and pseudo virtual state wavefunctions\/} \cite{os25,os28,os29}.
These seed solutions are obtained from the eigenfunctions by discrete symmetry operations (twisting)
\cite{os25,os29} or adopting the same function forms of the eigenfunctions with their degrees much higher
than the highest eigen level $n_\text{max}$ ({\em overshoot eigenfunctions\/}) \cite{os28}.
A Darboux transformation in terms of a virtual state wavefunction generates an iso-spectral deformation,
whereas a pseudo virtual state wavefunction used in a Darboux transformation will create an eigenstate
at its energy, which is below the original groundstate.
Therefore the latter generates a non-isospectral deformation.
The multi-index consists of the degrees of the used polynomial type seed solutions.
In this paper we simply use the same method (multiple Darboux transformations) and the same
polynomial type seed solutions to enlarge the list of exactly solvable scattering problems,
starting from five potentials having a finite number of eigenlevels, {\em e.g.} the Rosen-Morse 
potential, etc \cite{os28} and the Coulomb potential with the centrifugal barrier.
These six potentials are divided into three large groups, Group (A) for which the scattering waves extend
 from $x=-\infty$ to $x=+\infty$ and Group (B) for which the waves reach $x=+\infty$ 
 but they cannot reach $x=-\infty$, and Group (C) for the long-ranged Coulomb potential on half-line. 
The solvable full line scattering and half line scattering have very different characteristics.
In Group (A), the full line scattering cases, the transmission and reflection amplitudes are invariant
under the discrete symmetry transformation, whereas in Groups (B) and (C), the discrete symmetry of the potential 
does not imply   the invariance of the reflection amplitude.
The extended scattering data (the transmission and reflection amplitudes) are determined solely by the 
{\em asymptotic exponents\/} of the seed solutions, whereas the deformed potentials, in particular their 
regularity or non-singularity, depend heavily on the local behaviours of the seed solutions.

Historically, extensions of solvable scattering problems had been discussed in connection with
the soliton theory and the inverse scattering problem \cite{inv}. 
The adopted methods had been related to Abraham-Moses transformations \cite{A-M,os31} 
and the extensions were non-polynomial or non-rational \cite{AM-rel}--\cite{bay-sp}.
Compared to those non-rational extensions, the multi-indexed extensions reported in this paper
are much clearer in notion and
simpler in execution, although their applicability is limited to shape invariant potentials only.

The present paper is organised as follows. In section two the basic structures of the scattering problems 
and their extensions by multiple Darboux transformations are recapitulated.
The full line scattering, called Group (A) and the half line scattering, Groups (B) and (C), are treated separately.
Various properties of the polynomial type seed solutions, the virtual state, pseudo virtual state wavefunctions
and the overshoot eigenfunctions are explored in section three.
The conditions for the non-singularity of the deformed potentials are explained.
Sections four, five and six provides the explicit data of the original solvable potentials and the characteristics of the
deformed ones. 
The three potentials belonging to Group (A) are presented in section four: 
Rosen-Morse \S\ref{sec:RM}, soliton \S\ref{sec:sol} and hyperbolic symmetric top \S\ref{sec:hst} potentials.
Those belonging to Group (B) are explained in section five:  Morse \S\ref{sec:M}, Eckart \S\ref{sec:eck},
hyperbolic P\"oschl-Teller \S\ref{sec:hDPT}.  The long range Coulomb potential is the sole member of Group (C)
summarised in \S\ref{sec:coul} .
Section seven is for a summary and comments. 

\section{Scattering problems and their extensions}
\label{sec:scat}

Here we review the setting of one-dimensional quantum mechanical scattering problems and
their extensions by multiple Darboux transformations in terms of polynomial type seed solutions.
Let a quantum mechanical Hamiltonian $\mathcal{H}$ be defined  in an interval $x_1<x<x_2$, with a smooth potential:
\begin{equation}
\mathcal{H}=-\frac{d^2}{dx^2}+U(x),\quad U(x)\in\mathbb{R}.
  \label{schr}\\
\end{equation}
We discuss either the full line $x_1=-\infty$, $x_2=+\infty$ case called Group (A) or a half line 
$x_1=0$, $x_2=+\infty$ case called Group (B). 
Because of its long range character, Coulomb potential will be treated separately in (C). 
In all cases we adjust the constant part of the potential function so that it vanishes at the right infinity,
$U(+\infty)=0$.
We discuss the extensions of solvable potentials. That means the entire data of the discrete eigensystems 
and the scattering data are known:
\begin{align}
  &\mathcal{H}\phi_n(x)=\mathcal{E}_n\phi_n(x)
  \ \ (n=0,1,\ldots,n_{\text{max}}),\quad
  \mathcal{E}_0<\mathcal{E}_1<\cdots<\mathcal{E}_{n_{\text{max}}},
  \label{sheq}\\
  &(\phi_m,\phi_n)\eqdef\int_{x_1}^{x_2}\!dx\,\phi_m(x)\phi_n(x)
  =h_n\delta_{m\,n}\quad(h_n>0),
  \label{inpro}\\
&\mathcal{H}\psi_k(x)=k^2\psi_k(x),\qquad k\in\mathbb{R}_{\ge0},
\label{psik}\\
&
\psi_k(x)\approx\left\{
\begin{array}{cl}
 e^{ikx} &   \qquad x\to+\infty   \\
 A(k)e^{ikx}+B(k)e^{-ikx}&  \qquad  x\to-\infty 
\end{array}
\right.,\quad  (A),
\label{full}\\[4pt]
&
\psi_k(x)\approx \qquad r(k)e^{ikx}+e^{-ikx} \qquad  \qquad \ \    x\to+\infty ,
\quad \ \    (B),
\label{half}\\
&
\psi_k(x)\approx r(k)e^{ikx-i\gamma\log x}+e^{-ikx+i\gamma\log x} \quad     x\to+\infty ,
\quad  \  \   (C).
\label{coulasym}
\end{align}
Here $\gamma=-1/k$ for the Coulomb case.
The scattering amplitudes $A(k)$, $B(k)$ and $r(k)$   are {\em meromorphic\/}
functions of $k$, except for Rosen-Morse (RM)  \S\ref{sec:RM} and  Eckart \S\ref{sec:eck}, 
which are meromorphic functions of $k$ and $k'$ (to be defined there).
The transmission amplitude $t(k)$ and the reflection amplitude $r(k)$ for the full line scatterings are defined by:
\begin{align} 
  t(k)\eqdef\frac1{A(k)},\qquad r(k)&\eqdef\frac{B(k)}{A(k)},\qquad (A).
  \end{align}
In cases (B) and (C), the scattering wave $\psi_k(x)$  \eqref{half}--\eqref{coulasym} has to satisfy the square 
integrable boundary condition at the left boundary. 
In these cases, only the reflection amplitude $r(k)$ can be defined. In \eqref{coulasym} 
$\gamma$ is a coefficient of the logarithmic corrections due to the long range of Coulomb potential.
Except for the Coulomb potential \S\ref{sec:coul}, all the potentials discussed in this paper have
a finite number of discrete eigenlevels $n_{\text{max}}$.
As is well known that the poles on the {\em positive imaginary $k$-axis\/} of the transmission amplitude $t(k)$ 
((A) case)
and the reflection amplitude $r(k)$ ((B), (C) case) correspond to the {\em discrete eigenstates}.

The potentials to be discussed in this paper depend on a certain set of parameters, 
which is symbolically denoted by $\bm{\lambda}$.  When needed we express the parameter dependence
like $\mathcal{H}(\bm{\lambda})$, $U(x;\bm{\lambda})$, $t(k;\bm{\lambda})$, $r(k;\bm{\lambda}), etc$.
In Group (A), Rosen-Morse potential has two different asymptotic limits of the potential,
$U(-\infty)\neq U(+\infty)$.  This requires a slightly different setting from \eqref{full}.
It will be shown explicitly in \S\ref{sec:RM}.

We prepare polynomial type
seed solutions $\{\tilde{\phi}_{d_j}(x)\}$, $j=1,\ldots,M$, indexed by a set of non-negative integers
$\mathcal{D}=\{d_1,\ldots,d_M\}$ which are the {\em degrees\/} of the polynomial part of the seed solutions.
The asymptotic behaviours of the polynomial seed solution $\tilde{\phi}_\text{v}(x)$ are characterised by the
{\em asymptotic exponents\/} $\Delta_\text{v}^\pm$:
\begin{align} 
 \tilde{\phi}_\text{v}(x)&\approx
 \left\{
\begin{array}{cl}
e^{x\Delta_\text{v}^+} &  \qquad  x\to+\infty   \\
e^{x\Delta_\text{v}^-}&   \qquad x\to-\infty 
\end{array}
\right.,\quad (A) 
\label{asymfulls}\\
\tilde{\phi}_\text{v}(x)&\approx \quad \ e^{x\Delta_\text{v}^+}  \quad  \quad  \ \  x\to+\infty, \quad \ \  (B),\\
\tilde{\phi}_\text{v}(x)&\approx g_{\text{v}+}(x)\,e^{x\Delta_\text{v}^+}  \quad    x\to+\infty, \quad \ \  (C),
\label{asymcouls}
\end{align}
in which $g_{\text{v}+}(x)$ is a certain power function of $x$ due to the long range interactions.
See \eqref{coulasym2} for the explicit form.
The  extensions by multiple Darboux transformations in terms of polynomial type
seed solutions $\{\tilde{\phi}_{d_j}(x)\}$, $j=1,\ldots,M$ are expressed neatly in terms of a
ratio of Wronskians \cite{crum,os25,os28,os29}:
\begin{align}
  &\mathcal{H}^{[M]}\phi_{\mathcal{D},\,n}^{[M]}(x)=\mathcal{E}_n\phi_{\mathcal{D},\,n}^{[M]}(x),
  \quad (n=0,1,\ldots,n_{\text{max}}),
  \label{vDarb}\\
  &\phi_{\mathcal{D},\,n}^{[M]}(x)\eqdef
  \frac{\text{W}[\tilde{\phi}_{d_1},\tilde{\phi}_{d_2},\ldots,
  \tilde{\phi}_{d_M},\phi_{n}](x)}
  {\text{W}[\tilde{\phi}_{d_1},\tilde{\phi}_{d_2},\ldots,
  \tilde{\phi}_{d_M}](x)},\\
  &\mathcal{H}^{[M]}\psi_{\mathcal{D},\,k}^{[M]}(x)=k^2\psi_{\mathcal{D},\,k}^{[M]}(x),\quad k\in\mathbb{R}_{\ge0},
  \label{vDarb2}\\
  &\psi_{\mathcal{D},\,k}^{[M]}(x)\eqdef
  \frac{\text{W}[\tilde{\phi}_{d_1},\tilde{\phi}_{d_2},\ldots,
  \tilde{\phi}_{d_M},\psi_k](x)}
  {\text{W}[\tilde{\phi}_{d_1},\tilde{\phi}_{d_2},\ldots,
  \tilde{\phi}_{d_M}](x)},
  \label{Mpsidef}\\
  &\mathcal{H}^{[M]}\eqdef \mathcal{H}-2\partial_x^2\log
  \bigl|\text{W}[\tilde{\phi}_{d_1},\tilde{\phi}_{d_2},\ldots,
  \tilde{\phi}_{d_M}](x)\bigr|
  \label{vham}
\end{align}%
provided that the {\em deformed potential is non-singular\/}.  That requires the condition that  the Wronskian 
$\text{W}[\tilde{\phi}_{d_1},\tilde{\phi}_{d_2},\ldots, \tilde{\phi}_{d_M}](x)$
{\em should not have any zeros in the interval\/} $-\infty<x<\infty$ (A), or $0<x<\infty$ (B).
This condition will be discussed in \S\ref{sec:pol}.
For the discrete eigenstates $\{\phi_{\mathcal{D},\,n}^{[M]}(x)\}$ and the scattering states 
$\{\psi_{\mathcal{D},\,k}^{[M]}(x)\}$ the transformation is iso-spectral.
We stress, however, that additional discrete eigenstates may be created below the original groundstate level 
$\mathcal{E}_0$. Their number is equal to that of the used {\em pseudo virtual state wavefunctions\/}.

It should be stressed that with the polynomial type seed solutions \eqref{asymfulls}--\eqref{asymcouls},
the deformation potential $-2\partial_x^2\log
  \bigl|\text{W}[\tilde{\phi}_{d_1},\tilde{\phi}_{d_2},\ldots,
  \tilde{\phi}_{d_M}](x)\bigr|$ vanishes asymptotically, $x\to\pm\infty$.
Thus the deformed continuous spectrum also starts at $\mathcal{E}=0$ and the relationship between
the energy $\mathcal{E}$ and the wave number $k$, $\mathcal{E}=k^2$ is unchanged.
The multi-indexed scattering amplitudes are easily obtained from the asymptotic form of the
wavefunction $\psi_{\mathcal{D},\,k}^{[M]}(x)$ \eqref{Mpsidef} by using the asymptotic forms of the original 
wavefunction $\psi_k(x)$ \eqref{full}--\eqref{coulasym} and those of the polynomial seed solutions 
$\tilde{\phi}_\text{v}(x)$ \eqref{asymfulls}--\eqref{asymcouls}.
For the full line scattering (Group (A)) case, we obtain
\begin{align} 
\psi_{\mathcal{D},\,k}^{[M]}(x)&\approx \prod_{j=1}^M (ik-\Delta_{d_j}^+)\cdot e^{ikx} 
\hspace{71mm} x\to+\infty, \\
 \psi_{\mathcal{D},\,k}^{[M]}(x)&\approx \prod_{j=1}^M (ik-\Delta_{d_j}^-)\cdot A(k)\,e^{ikx}
 + \prod_{j=1}^M (-ik-\Delta_{d_j}^-)\cdot B(k)\,e^{-ikx}\qquad x\to-\infty,  
 \label{fullleft}
\end{align}
which lead to multi-indexed multiplicative deformations of the transmission and reflection amplitudes:
\begin{equation}
\text{(A)}:\quad
t_{\mathcal D}(k)=\prod_{j=1}^M\frac{k+i\Delta_{d_j}^+}{k+i\Delta_{d_j}^-}\cdot t(k),\quad
r_{\mathcal D}(k)=(-1)^M\prod_{j=1}^M\frac{k-i\Delta_{d_j}^-}{k+i\Delta_{d_j}^-}\cdot r(k).
\label{trDfull}
\end{equation}
For Rosen-Morse potential (see \S\ref{sec:RM} for the definition of $k'$), we obtain 
\begin{equation}
\text{RM}:\quad
t_{\mathcal D}(k)=\prod_{j=1}^M\frac{k'+i\Delta_{d_j}^+}{k+i\Delta_{d_j}^-}\cdot t(k),\quad
r_{\mathcal D}(k)=(-1)^M\prod_{j=1}^M\frac{k-i\Delta_{d_j}^-}{k+i\Delta_{d_j}^-}\cdot r(k).
\label{trDRM}
\end{equation}
For the half line scattering (Group (B), (C)) case, similar calculation gives
\begin{align} 
& \text{(B)}:\quad \psi_{\mathcal{D},\,k}^{[M]}(x)\approx \prod_{j=1}^M (ik-\Delta_{d_j}^+)\cdot r(k)\,e^{ikx}
 + \prod_{j=1}^M (-ik-\Delta_{d_j}^+)\cdot\,e^{-ikx}\qquad x\to+\infty,  
 \label{defhalf}\\
 & \text{(C)}:\quad  \psi_{\mathcal{D},\,k}^{[M]}(x)\approx \prod_{j=1}^M (ik-\Delta_{d_j}^+)\cdot r(k)\,e^{ikx-i\gamma \log x}
 + \prod_{j=1}^M (-ik-\Delta_{d_j}^+)\cdot\,e^{-ikx+i\gamma \log x}
 \quad \  x\to+\infty,  
 \label{defhalfcoul}\\
 &\qquad  \text{(B), (C)}:\quad
r_{\mathcal D}(k)=(-1)^M\prod_{j=1}^M\frac{k+i\Delta_{d_j}^+}{k-i\Delta_{d_j}^+}\cdot r(k).
\label{rBC}
\end{align}
It should be stressed that the meromorphic character of the scattering amplitudes is preserved by the
multi-indexed extensions and that the added poles and zeros all appear on the imaginary $k$-axis 
determined solely by the asymptotic exponents of the used polynomial type seed solutions.
The derivation depends on the simple fact:
the Wronskian of exponential functions $\text{W}[e^{\alpha_1x},e^{\alpha_2x},\ldots,e^{\alpha_M x}](x)$
is  reduced to a van der Monde determinant
\begin{equation*}
\text{W}[e^{\alpha_1x},e^{\alpha_2x},\ldots,e^{\alpha_M x}](x)=
\prod_{1\le k<j\le M}(\alpha_j-\alpha_k)\cdot e^{\sum_{j=1}^M \alpha_j x},
\end{equation*}
and most factors cancel out between the numerator and denominator of \eqref{Mpsidef}.
The logarithmic \eqref{coulasym} and power function \eqref{asymcouls} correction terms of 
Coulomb case give vanishing asymptotic contributions ($\partial_x\log x=1/x\to 0$) in the Wronskian,  and
the final deformation formula  is the same \eqref{rBC} as that of the other half line scattering potentials.

\section{Polynomial type seed solutions}
\label{sec:pol}

In the previous section we have derived simple formulas of the transmission  and
the reflection  amplitudes \eqref{trDfull}--\eqref{rBC} after the multi-indexed extensions.
The deformed amplitudes $t_\mathcal{D}(k)$ and $r_\mathcal{D}(k)$ are again meromorphic functions of
$k$ with $M$ extra zeros and poles  determined by the asymptotic exponents $\Delta_{d_j}^\pm$ 
appearing on the imaginary axis. However, this does not necessarily mean that any extension 
specified by $\mathcal{D}=\{d_1,\ldots,d_M\}$ are realisable.
The deformed potential \eqref{vham} must be non-singular, that is,  the Wronskian 
$\text{W}[\tilde{\phi}_{d_1},\tilde{\phi}_{d_2},\ldots, \tilde{\phi}_{d_M}](x)$
 should not have any zeros in the interval where the scattering takes place.
The situation is in good contrast with the extensions by Abraham-Moses transformations \cite{A-M,os31},
in which case the non-singularity of the deformed potential is always guaranteed 
but simple formulas of the deformed amplitudes are not available due to the non-rational nature of the extensions.

In this section we summarise the properties of various polynomial type seed solutions,
virtual and pseudo virtual state  wavefunctions and overshoot eigenfunctions and provide several practical rules
for achieving non-singular extensions.
Most of the materials in this section have been presented in connection with rational extensions of
solvable potentials, in particular, the bound state problems, \cite{os25,os28,os29}.
The well informed readers can skip this section.

First of all, these polynomial seed solutions $\{\tilde{\phi}_\text{v}(x)\}$ are required to have 
{\em energies below the original groundstate energy\/} $\tilde{\mathcal E}_\text{v}<\mathcal{E}_0$.
This is enough to guarantee the positive norm of all the deformed eigenstates:
\begin{align}
  (\phi_{\mathcal{D},\,m}^{[M]},\phi_{\mathcal{D},\,n}^{[M]})
  =\prod_{j=1}^M(\mathcal{E}_n-\tilde{\mathcal{E}}_{d_j})\cdot
  h_n\delta_{m\,n}, \quad (n,m=0,1,\ldots,n_{\text{max}}).
  \label{Mnorm2}
\end{align}
However this is not a sufficient condition for the non-singularity of the deformed potential.
But it is closely related with the absence of zeros of the seed solutions  in the physical region.

A {\em virtual state wavefunctions} 
$\tilde{\phi}_\text{v}(x)$ is obtained from the eigenfunction {\em by a discrete symmetry transformation\/},
namely by a certain twist of parameters.
Thus they are {\em non-square integrable polynomial type solutions\/} of the original Schr\"odinger 
equation \eqref{schr},
 $\mathcal{H}\tilde{\phi}_\text{v}(x)
=\tilde{\mathcal{E}}_\text{v}\tilde{\phi}_\text{v}(x)$:
The degree $\text{v}$ of the polynomial is so restricted that the corresponding energy is below the original 
groundstate level $\tilde{\mathcal{E}}_\text{v}<\mathcal{E}_0$ and it has 
{\em no zeros in the domain\/} $x_1<x<x_2$.
Thus a Darboux transformation in terms of a virtual state wavefunction is non-singular.

There are two types of virtual state wavefunctions, type I and type II.
The type I virtual state wavefunctions $\{\tilde{\phi}_\text{v}^\I\}$ are square non-integrable 
at the upper boundary $x_2$ and their reciprocals are square non-integrable at the lower boundary $x_1$.
That is $\{\tilde{\phi}_\text{v}^\I\}$ go to zero fairly rapidly as $x\to x_1$.
For type II the behavior is opposite, $\{\tilde{\phi}_\text{v}^\II\}$ are square non-integrable 
at the lower boundary $x_1$ and their reciprocals are square non-integrable at the upper boundary $x_2$.
Since a Wronskian of two type I (II) virtual state wavefunctions  is monotonously increasing or decreasing
\begin{equation*}
\partial_x\text{W}[\tilde{\phi}_{\text{v}_1}^\I,\tilde{\phi}_{\text{v}_2}^\I](x)=
(\tilde{\mathcal E}_{\text{v}_1}-\tilde{\mathcal E}_{\text{v}_2})\tilde{\phi}_{\text{v}_1}^\I(x)\tilde{\phi}_{\text{v}_2}^\I(x)
\ppm0,
\end{equation*}
so the boundary condition of vanishing at the lower boundary  is ``preserved". 
Thus the above Wronskian has also no zeros in the domain.
Multiple Darboux transformations in terms of type I (II) virtual state wavefunctions only, 
$\text{W}[\tilde{\phi}^\I_{d_1},\tilde{\phi}^\I_{d_2},\ldots,\tilde{\phi}^\I_{d_M}](x)$
is {\em non-singular\/}
provided the parameter ranges are properly restricted such that the conditions
\begin{equation}
 \partial_x^s\tilde{\phi}^\I_{d_j}(x)\bigm|_{x=\text{$x_1$}}=0
 \ \ (s=0,1,\ldots,M-1)
 \label{multicond}
\end{equation}
are satisfied. 
For the  Darboux transformations in terms of mixed type I and II virtual state wavefunctions, 
proper care is needed to ensure the non-singularity of the resulting potentials. 

The {\em pseudo virtual state wavefunctions\/} are also obtained from the eigenfunctions 
by discrete symmetry transformations, {\em i.e.\/} by twists of the parameters.
Thus they are also polynomial type solutions specified by their degrees $\{\text{v}\}$. 
In this case the requirements on the degrees are that they are square non-integrable at both 
boundaries and their reciprocals are square integrable at both boundaries.
Another requirement is that their energies satisfy the condition
\begin{equation}
\tilde{\mathcal E}_\text{v}=\mathcal{E}_{-(\text{v}+1)}.
\label{psener}
\end{equation}
The nodeless condition on each particular pseudo virtual state wavefunction is not required.

The conditions of non-singularity for multiple Darboux transformations in terms of pseudo virtual wavefunctions
specified by their degrees are stated through their equivalent  {\em eigenstates deleting\/} Darboux transformations
\cite{os29,adler}.
Let $\mathcal{D}\eqdef\{d_1,d_2,\ldots,d_M\}$ ($d_j\in\mathbb{Z}_{\ge0}$)
be a set of distinct non-negative integers.
We introduce an integer $N$ as the maximum of
$\mathcal{D}$, $N\eqdef\text{max}(\mathcal{D})$.
Let us define a set of distinct non-negative integers
$\bar{\mathcal{D}}=\{0,1,\ldots,N\}\backslash
\{\bar{d}_1,\bar{d}_2,\ldots,\bar{d}_M\}$
together with the shifted parameters $\bar{\bm{\lambda}}$:
\begin{align}
  &\bar{\mathcal{D}}\eqdef\{0,1,\ldots,\breve{\bar{d}}_1,\ldots,
  \breve{\bar{d}}_2,\ldots,\breve{\bar{d}}_M,\ldots,N\}
  =\{e_1,e_2,\ldots,e_{N+1-M}\},\\
  &\bar{d}_j\eqdef N-d_j,\quad
  \bar{\bm{\lambda}}\eqdef \bm{\lambda}-(N+1)\bm{\delta}.
  \label{barD}
\end{align}
Here $\bm{\delta}$ is a characteristic quantity of each solvable potential, called the `shifts', whose values are 
given in sections four, five and six for each solvable potential. 
The notation $\breve{\bar{d}}_j$ simply means that $\bar{d}_j$ should be excluded from the containing set.
The singularity free conditions of the $M$-th deformed potential \eqref{vham} are \cite{adler,os29}:
\begin{equation}
  \prod_{j=1}^{N+1-M}(n-e_j)\ge0
  \quad(\,\forall n\in\mathbb{Z}_{\geq 0}).
  \label{non-sing}
\end{equation}
For $M=1$, $\mathcal{D}=\{d_1\}$,
$\bar{\mathcal{D}}=\{1,2,\ldots,d_1\}$,
the above conditions are satisfied by even $d_1$,
$d_1\in 2\mathbb{Z}_{\geq 0}$.
For example, $\mathcal{D}=\{2,3\}$,  $\{2,5\}$, $\{4,5\}$,  $\{2,3,4\}$, $\{2,5,6\}$, $\{2,5,8\}$, $\{2,5,8,11\}$ 
etc produce non-singular potentials.

The {\em overshoot eigenfunctions\/} are another category of polynomial type seed solutions \cite{os28}.
The overshoot eigenfunctions have exactly the same forms as the eigenfunctions
$\tilde{\phi}^{\text{os}}_{\text{v}}(x)\eqdef\phi_{\text{v}}(x)$
($\tilde{\mathcal{E}}^{\text{os}}_{\text{v}}=\mathcal{E}_{\text{v}}$).
But their degrees are much higher than the highest discrete energy level
$n_{\text{max}}$ so that their energies are lower than the groundstate energy $\mathcal{E}_0$.
Depending on their asymptotic behaviours, the overshoot eigenfunctions are classified into
type I  and type II virtual state wavefunctions and `pseudo virtual' state wavefunctions, which fail to satisfy the
condition \eqref{psener}, $\tilde{\mathcal{E}}^{\text{os}}_{\text{v}}\neq\mathcal{E}_{-\text{v}-1}$.
For some potentials, the classification of the seed solutions changes depending on their degrees $\text{v}$.
As for the `pseudo virtual' wavefunctions obtained from the overshoot eigenfunctions, 
the above criterion for non-singularity \eqref{non-sing} cannot be used, since \eqref{psener} is not satisfied.

When various kinds of polynomial type seed solutions, type I and II virtual, pseudo virtual and `pseudo virtual',
are  employed for multiple Darboux transformations, the non-singularity of the deformed potentials should
be verified one by one, as we do  not have as yet a general criterion.

The asymptotic behaviours of the polynomial type seed solutions are summarised neatly as the signs of the corresponding asymptotic exponents:
\begin{align*} 
\text{type I virtual:}& \quad \Delta_\text{v}^+>0,\quad  \Delta_\text{v}^->0,\quad 
\Delta_\text{v}^+\Delta_\text{v}^->0,\\
\text{type II virtual:}& \quad \Delta_\text{v}^+<0,\quad  \Delta_\text{v}^-<0,\quad 
\Delta_\text{v}^+\Delta_\text{v}^->0,\\
\text{pseudo virtual:}& \quad \Delta_\text{v}^+>0,\quad  \Delta_\text{v}^-<0,\quad 
\Delta_\text{v}^+\Delta_\text{v}^-<0.
\end{align*}
For the half line scattering, only the first entries are meaningful.

As explained above, type I and II virtual state wavefunctions generate {\em iso-spectral\/} deformations.
That means, no new eigenstates are created.
As seen from \eqref{trDfull}, \eqref{trDRM}, a type II virtual state ($\Delta_\text{v}^-<0$) contributes a new 
pole factor of the transmission amplitude  $t_{\mathcal D}(k)$ {\em on the positive imaginary $k$-axis\/}.
Among  the three potentials of Group (A), only Rosen-Morse potential has type II virtual state wavefunctions.
Close examination in \S\ref{sec:RM} reveals that the numerator factor $k'+i\Delta_\text{v}^+$ cancels that pole.
As for the half line scatterings, a type I virtual state ($\Delta_\text{v}^+>0$) contributes 
a new pole factor of the reflection
amplitude $r(k)$ on the positive imaginary $k$-axis.
Among the half line scattering potentials, Morse \S\ref{sec:M}, Eckart \S\ref{sec:eck} and 
hyperbolic PT \S\ref{sec:hDPT} potentials have type I virtual states. 
As shown for each potential in \S\ref{sec:half}, these new poles at $k=i\Delta_\text{v}^+$
are cancelled by the zeros of the scattering amplitude $r(k)$.
In contrast, a pseudo or a  `pseudo' virtual state wavefunction, having $\Delta_\text{v}^->0$
for full line and $\Delta_\text{v}^+>0$ for half line, {\em always creates a new pole\/} 
on the positive imaginary $k$-axis of $t(k)$ (full) and $r(k)$ (half).
These new poles correspond to the energy of the added eigenstates.
For each potential in sections four, five and six, we demonstrate quite simply that the energies of the created
eigenstates are equal to the energies of the employed pseudo and `pseudo' virtual state wavefunctions
in accordance with the general theory of Darboux transformations \cite{os29}.

\goodbreak


\section{Original and Extended Scatterings: \\
Full line scattering}
\label{sec:full}

In the  three sections hereafter, we first present various data of the original solvable potentials;
the eigenenergies, eigenfunctions, scattering data, {\em i.e,\/} the transmission and reflection amplitudes,
various polynomial seed solutions, the virtual and pseudo virtual state wavefunctions, 
the overshoot eigenfunctions together with the corresponding asymptotic exponents. 
These original solvable potentials are all {\em shape invariant\/} \cite{genden}.
We emphasize that the original scattering amplitudes presented in this article satisfy the constraints of shape invariance:
\begin{align} 
\text{full line:}\quad t(k;\bm{\lambda}+\bm{\delta})&=\left(\frac{ik+W_+}{ik+W_-}\right) t(k;\bm{\lambda}),\quad
r(k;\bm{\lambda}+\bm{\delta})=\left(\frac{-ik+W_-}{ik+W_-}\right) r(k;\bm{\lambda}),\\
\text{RM}:\quad
t(k;\bm{\lambda}+\bm{\delta})&=\left(\frac{ik'+W_+}{ik+W_-}\right) t(k;\bm{\lambda}),\quad
r(k;\bm{\lambda}+\bm{\delta})=\left(\frac{-ik+W_-}{ik+W_-}\right) r(k;\bm{\lambda}),\\
\text{half line}:\quad r(k;\bm{\lambda}+\bm{\delta})&=\left(\frac{ik+W_+}{-ik+W_+}\right) r(k;\bm{\lambda}),
\label{halfshape}\\
W_+&\eqdef-\lim_{x\to+\infty}\frac{\partial_x\phi_0(x;\bm{\lambda})}{\phi_0(x;\bm{\lambda})},\qquad
W_-\eqdef-\lim_{x\to-\infty}\frac{\partial_x\phi_0(x;\bm{\lambda})}{\phi_0(x;\bm{\lambda})}.
\end{align}
For simplicity of presentation, we do not indicate the parameter dependence of various exponents, 
$\Delta^\pm_\text{v}$, $W_\pm$, etc. But of course they depend on the parameters.
The characteristic properties of  various extended potentials and the deformed scattering data are
discussed.
For various properties of virtual and pseudo virtual wavefunctions and overshoot eigenfunctions
including their applications
we refer to \cite{os28,os29} for more details.
The solvable potentials are divided into three groups: the full line scattering  in Group (A),
the half line scattering in Group (B) and the long range scattering in Group (C).

This Group (A) consists of three potentials: Rosen-Morse (RM), soliton (s) and hyperbolic symmetric top II (hst).

\subsection{Rosen-Morse potential (RM)}
\label{sec:RM}

\paragraph{Original system}

This potential is also called Rosen-Morse $\II$ potential.
The system has finitely many discrete eigenstates
$0\le n\le n_\text{max}(\bm{\lambda})=[h-\sqrt{\mu}\,]'$ in the specified
parameter range ($[a]'$ denotes the greatest integer not exceeding and not equal to $a$):
\begin{align*}
  &\bm{\lambda}=(h,\mu),\quad\bm{\delta}=(-1,0),\quad
  -\infty<x<\infty,\quad h(h-1)>\mu>0,\\
  &  U(x;\bm{\lambda})=-\frac{h(h+1)}{\cosh^2x}+2\mu\left(\tanh x +1\right)
 ,\\
  &\mathcal{E}_n(\bm{\lambda})=
 -\left[h-n-\frac{\mu}{(h-n)}\right]^2,\quad\eta(x)=\tanh x,\n\\
  &\phi_n(x;\bm{\lambda})=e^{-\frac{\mu}{h-n}x}(\cosh x)^{-h+n}
  P_n^{(\alpha_n,\beta_n)}(\eta),\quad W_+=\alpha_0,\quad W_-=-\beta_0.\\ 
  &
  \alpha_n=h-n+\frac{\mu}{h-n},\ \ \beta_n=h-n-\frac{\mu}{h-n}.
\end{align*}
The asymptotic behaviour of this potential is slightly different from the others in the full line scattering.
 Although it approaches to zero at $-\infty$, $U(-\infty;\bm{\lambda})\to 0$, in the limit $x\to  \infty$ 
 it approaches a non-vanishing constant, $U(+\infty;\bm{\lambda})\to U_0 \equiv 4\mu>0$.
The asymptotic forms of the scattering state are slightly different from \eqref{full}
\begin{align}
\psi_k(x;\bm{\lambda})\sim
\left\{
  \begin{array}{cl}
 e^{ik^\prime x}
  & \qquad x\to +\infty, \\
  A(k;\bm{\lambda})  e^{ik x} + B(k;\bm{\lambda}) e^{-ik x} & \qquad x \to -\infty,
  \end{array}\right.
\end{align}
in which $k^\prime\eqdef\sqrt{k^2-4\mu}$ is the wave number in the region   $x>0$.
We choose the Riemann sheet such that $k'\to k$ in the limit $\mu\to0$.
 The amplitudes $A(k;\bm{\lambda})$ and $B(k;\bm{\lambda})$ are given by \cite{KS}
\begin{align}
A(k;\bm{\lambda}) &= \frac{\Gamma(-ik)\Gamma(1-ik^\prime )}{\Gamma(-h-i\frac{k}{2}-i\frac{k^\prime}{2})\,
\Gamma(1+h-i\frac{k}{2}-i\frac{k^\prime}{2})},
\label{ampa}\\
B(k;\bm{\lambda}) &=  \frac{\Gamma(ik)\Gamma(1-ik^\prime)}{\Gamma(-h+i\frac{k}{2}-i\frac{k^\prime}{2})\, 
\Gamma(1+h+i\frac{k}{2}-i\frac{k^\prime}{2})}.
\label{amp}
\end{align}
These give
\begin{align}
t(k;\bm{\lambda})&=\frac{\Gamma(-h-i\frac{k}{2}-i\frac{k^\prime}{2})\Gamma(1+h-i\frac{k}{2}-i\frac{k^\prime}{2})}
{\Gamma(-ik)\Gamma(1-ik^\prime )},
\label{RMt}\\
r(k;\bm{\lambda}) &=  \frac{\Gamma(ik)\Gamma(-h-i\frac{k}{2}-i\frac{k^\prime}{2})
\Gamma(1+h-i\frac{k}{2}-i\frac{k^\prime}{2})}
{\Gamma(-ik)\Gamma(-h+i\frac{k}{2}-i\frac{k^\prime}{2})
\Gamma(1+h+i\frac{k}{2}-i\frac{k^\prime}{2})}.
\label{RMr}
\end{align}
The poles on the positive imaginary $k$-axis coming from the first Gamma function factor in the numerator of 
$t(k;\bm{\lambda})$ \eqref{RMt}, $-h-i\frac{k}2-i\frac{k'}2=-n$, $\Rightarrow k=i\beta_n$, $\beta_n>0$, 
$n=0,1,\ldots, [h-\sqrt{\mu}]'$ provide the eigenspectrum as above.
In the $\mu\to0$ limit, the potential, eigenvalues, eigenfunctions, scattering amplitudes and the polynomial type 
seed solutions  give the corresponding quantities of  soliton potential.

The physical quantities for the one-dimensional scattering problem  are the  transmission ($T(k;\bm{\lambda})$) and reflection 
($R(k;\bm{\lambda})$) coefficients, defined as the ratio of the fluxes of the transmitted and reflected waves, 
respectively, to that of the incident wave.  They are related to $t(k;\bm{\lambda})$ and $r(k;\bm{\lambda})$ by
\begin{equation}
T(k;\bm{\lambda})=\frac{k^\prime}{k}\,|t(k;\bm{\lambda})|^2 =
 \frac{k^\prime}{k}\cdot\frac{1}{|A(k;\bm{\lambda})|^2},\quad
 R(k;\bm{\lambda})=|r(k;\bm{\lambda})|^2=\left|\frac{B(k;\bm{\lambda})}{A(k;\bm{\lambda})}\right|^2,\quad k\in\mathbb{R}_{\ge0}.
\label{TRdef}
\end{equation}
It is easy to verify   $T(k;\bm{\lambda})+R(k;\bm{\lambda})=1$, which indicates the conservation of total flux.
Note that, for energy $0<k^2<U_0$, the wave number $k^\prime$ becomes purely imaginary. 
 In this case, there is no transmitted waves, i.e., $T(k;\bm{\lambda})=0$ and $R(k;\bm{\lambda})=1$. 
Setting $k^\prime=k$ gives the corresponding expressions for the case of soliton potential to be discussed 
in the next subsection.

The potential $U(x;h,\mu)$ and the scattering data \eqref{ampa}--\eqref{RMr} are invariant under the discrete
transformation $h\to -(h+1)$, $\mu\to \mu$. But the eigenvalues and the eigenfunctions are not invariant.
The discrete symmetry transformations produce polynomial type seed solutions, as shown below.

\paragraph{Polynomial type seed solutions}

The discrete symmetry $h\to -h-1$, $\mu\to\mu$ generates the pseudo virtual state wavefunctions:
\begin{align}
  &\text{pseudo virtual}:\quad \tilde{\phi}_{\text{v}}(x;\bm{\lambda})
  =e^{\frac{\mu}{h+1+\text{v}}x}(\cosh x)^{h+1+\text{v}}
  P_{\text{v}}^{(\bar{\alpha}_\text{v},\bar{\beta}_\text{v})}(\eta)
  \quad(\text{v}\in\mathbb{Z}_{\ge0}),\n\\
&\Delta_{\text{v}}^+=-\bar{\alpha}_\text{v}>0,\quad
\Delta_{\text{v}}^-=\bar{\beta}_\text{v}<0,\qquad \tilde{\mathcal{E}}_{\text{v}}(\bm{\lambda})
    =\mathcal{E}_{-\text{v}-1}(\bm{\lambda})=-\bar{\beta}_\text{v}^2,
  \label{RMpv}\\
  &
 \bar{\alpha}_\text{v}=-h-1-\text{v}-\frac{\mu}{h+1+\text{v}},\ \ 
 \bar{\beta}_\text{v}=-h-1-\text{v}+\frac{\mu}{h+1+\text{v}}.
 \label{RMpv2}
\end{align}
For non-negative $\text{v}$ the above pseudo virtual state wavefunctions have lower energies than 
the groundstate
energy $\tilde{\mathcal{E}}_{\text{v}}(\bm{\lambda})<\mathcal{E}_0(\bm{\lambda})$.

The overshoot eigenfunctions provide two types of virtual state wavefunctions
\cite{quesne5} and `pseudo virtual' state  wavefunctions \cite{os28}:
\begin{align}
  &\tilde{\phi}^{\text{os}}_{\text{v}}(x;\bm{\lambda})
  =\phi_{\text{v}}(x;\bm{\lambda}),\qquad \tilde{\mathcal E}_{\text{v}}(\bm{\lambda})=-{\beta}_\text{v}^2,\n\\
  &
\Delta_{\text{v}}^+=-{\alpha}_\text{v},\quad \Delta_{\text{v}}^-={\beta}_\text{v},\ 
  \left\{\begin{array}{rlll}
  \text{type $\II$ virtual}:&
  {\displaystyle h-\frac{\mu}{h}<\text{v}<h},&\Delta_{\text{v}}^+<0, &\Delta_{\text{v}}^-<0\\
  \text{type $\I$ \ virtual}:&
  {\displaystyle h<\text{v}<h+\frac{\mu}{h}},&\Delta_{\text{v}}^+>0, &\Delta_{\text{v}}^->0\\
  \text{`pseudo virtual'}:&
  {\displaystyle \text{v}>2h},&\Delta_{\text{v}}^+>0, &\Delta_{\text{v}}^-<0
  \end{array}\right..
  \label{RMover}
\end{align}
It is obvious that the above type I and II virtual states disappear in the soliton potential, {\em i.e.} $\mu\to0$ limit.

\subsubsection{Deformed scatterings}
Here we examine various properties of the multi-indexed extended scattering amplitudes \eqref{trDRM} 
derived in section two.

\paragraph{Invariant transmission and reflection coefficients}

For real $k\in\mathbb{R}_{\ge0}$, the  additional factors in \eqref{trDRM} 
$\prod_{j=1}^M(k'+i\Delta_{d_j}^+)/(k+i\Delta_{d_j}^-)$, $\prod_{j=1}^M(k-i\Delta_{d_j}^-)/(k+i\Delta_{d_j}^-)$,
are all modulo one. We conclude that the transmission  and reflection coefficients \eqref{TRdef} are the same as 
those of the original potentials:
\begin{equation}
T_{\mathcal D}(k;\bm{\lambda})=T(k;\bm{\lambda}),\quad R_{\mathcal D}(k;\bm{\lambda})=R(k;\bm{\lambda}).
\end{equation}
These  relations are a generic feature of the multi-indexed extensions and they  hold for all the examples 
discussed in this paper.

\paragraph{Type II virtual}

Here we examine the apparent poles on the positive imaginary $k$-axis of the transmission 
amplitude $t_{\mathcal D}(k)$ caused by the overshoot 
type II virtual state wavefunctions \eqref{RMover} having $\Delta_\text{v}^->0$. In \eqref{trDRM} 
the contribution of a type II virtual state wavefunction $\tilde{\phi}_\text{v}(x)$ reads:
\begin{equation*}
\frac{k'+i\Delta_\text{v}^+}{k+i\Delta_\text{v}^-}=\frac{k'-i\alpha_\text{v}}{k+i\beta_\text{v}}
=\frac{k-i\beta_\text{v}}{k'+i\alpha_\text{v}}\rightarrow-\frac{\beta_\text{v}}{\alpha_\text{v}}, \quad
 \text{as} \quad {k\to -i\beta_{\text{v}}}.
\end{equation*}
As $k$ approaches $-i\beta_\text{v}$, the numerator also vanishes and the ratio goes to a constant.
This is a simple consequence of the phase convention of $k'$ and $k'^2+\alpha_\text{v}^2=k^2+\beta_\text{v}^2$.
On the other hand, type I virtual state wavefunctions produce poles in $t_{\mathcal D}(k;\bm{\lambda})$ and 
$r_{\mathcal D}(k;\bm{\lambda})$ on the negative imaginary $k$-axis. 
As for the poles in $t_{\mathcal D}(k;\bm{\lambda})$, 
the same cancellation mechanism as above works and the pole disappears.
As for the poles in $r_{\mathcal D}(k;\bm{\lambda})$, the second Gamma function factor in the denominator cancels them
for both type I, II and pseudo virtual:
\begin{equation*}
\frac1{\Gamma(-h+i\frac{k}{2}-i\frac{k^\prime}{2})}\propto \frac12\left(-h+\text{v}+\frac{\mu}{h-\text{v}}+ik\right)
=\frac12\left(ik-\beta_\text{v}\right), \quad h-\frac{\mu}h<\text{v}.
\end{equation*}

\paragraph{Pseudo virtual}
A pseudo (`pseudo') virtual wavefunction will add a new discrete eigenstate at its energy.
It is trivial to verify that the pseudo \eqref{RMpv} and `pseudo' \eqref{RMpv2} virtual will add a pole 
on the positive imaginary $k$-axis with exactly the same energy of the employed seed solution,
$-\bar{\beta}_\text{v}^2$ and $-\beta_\text{v}^2$, respectively.


\subsection{Soliton potential}
\label{sec:sol}

\paragraph{Original system}
In the limit $\mu\to 0$, the RM potential reduces to  soliton potential. 
The system has finitely many discrete eigenstates
$0\le n\le n_\text{max}(\bm{\lambda})=[h]'$ in the specified parameter range:
\begin{align*}
  &\bm{\lambda}=h,\quad \bm{\delta}=-1,\quad
  -\infty<x<\infty,\quad h>1/2,\\
  &
  U(x;\bm{\lambda})=-\frac{h(h+1)}{\cosh^2x},\\
  &\mathcal{E}_n(\bm{\lambda})=-(h-n)^2,\quad \eta(x)=\tanh x,\\
  &\phi_n(x;\bm{\lambda})=(\cosh x)^{-h+n}\times P^{(h-n,h-n)}_n(\tanh x),\quad W_+=-W_-=h.
\end{align*}
The transmission and reflection amplitudes are:
\begin{align}
t(k;\bm{\lambda})&=\frac{\Gamma(-h-ik)\Gamma(1+h-ik)}{\Gamma(-ik)\Gamma(1-ik )},
\label{solt}\\
r(k;\bm{\lambda}) &=  \frac{\Gamma(ik)\Gamma(-h-ik)\Gamma(1+h-ik)}{\Gamma(-ik)\Gamma(-h)
\Gamma(1+h)}.
\label{solr}
\end{align}
The poles on the positive imaginary $k$-axis coming from the first Gamma function factor in the numerator 
of $t(k;\bm{\lambda})$ \eqref{solt}, $-h-i{k}=-n$, $\Rightarrow k=i(h-n)$,  $n=0,1,\ldots, [h]'$ provide the eigenspectrum 
as above. As is well known, at integer $h\in\mathbb{Z}_{\ge1}$ the reflectionless potential 
$r(k;\bm{\lambda})=0$ is realised by the poles of the Gamma function $\Gamma(-h)$ 
in the denominator of $r(k;\bm{\lambda})$.
The potential and the scattering amplitudes \eqref{solt}--\eqref{solr} are invariant under the
discrete transformation $h\to -(h+1)$, but the eigenvalues and the eigenfunctions are not.
The relation $|t(k;\bm{\lambda})|^2+|r(k;\bm{\lambda})|^2=1$ holds as the $\mu\to0$ limit of Rosen-Morse case.

\paragraph{Polynomial type seed solutions}

The discrete symmetry $h\to-h-1$ generates the pseudo virtual  wavefunctions, which lie below the groundstate:
\begin{align}
  &\text{pseudo virtual}:\quad \tilde{\phi}_{\text{v}}(x;\bm{\lambda})
  =(\cosh x)^{h+1+\text{v}}P_{\text{v}}^{(-h-1-\text{v},-h-1-\text{v})}(\tanh x)
  \quad(\text{v}\in\mathbb{Z}_{\ge0}),\n\\
&\Delta_{\text{v}}^+=h+1+\text{v}>0,\quad 
\Delta_{\text{v}}^-=-\Delta_{\text{v}}^+<0, \quad
  \tilde{\mathcal{E}}_{\text{v}}(\bm{\lambda})
    =\mathcal{E}_{-\text{v}-1}(\bm{\lambda})<\mathcal{E}_0(\bm{\lambda}).
    \label{solpv}
\end{align}
The overshoot eigenfunctions provide `pseudo' virtual state  wavefunctions for this potential
 for $\text{v}>2h$:
\begin{align}
  &\text{`pseudo virtual'}:\quad \tilde{\phi}^{\text{os}}_{\text{v}}(x;\bm{\lambda})
  =\phi_{\text{v}}(x;\bm{\lambda}),\n\\
&\Delta_{\text{v}}^+=-h+\text{v}>0,\quad 
\Delta_{\text{v}}^-=-\Delta_{\text{v}}^+<0,
\quad
  \tilde{\mathcal{E}}^{\text{os}}_{\text{v}}(\bm{\lambda})
  <\mathcal{E}_0(\bm{\lambda}) \quad(\text{v}>2h).
      \label{solpv2}
\end{align}

\paragraph{Deformed scatterings}
A pseudo (`pseudo') virtual state wavefunction will add a new discrete eigenstate at its energy.
It is trivial to verify that the pseudo \eqref{solpv} and `pseudo' \eqref{solpv2} virtual wavefunctions will add a pole 
on the positive imaginary $k$-axis at $k=i(\text{v}+h+1)$, $k=i(\text{v}-h)$, respectively,
 with exactly the same energy of the employed seed solution,
$-(h+\text{v}+1)^2$ and $-(\text{v}-h)^2$, respectively.
For both the pseudo \eqref{solpv} and `pseudo' \eqref{solpv2} virtual wavefunctions, 
$\Delta_\text{v}^+=-\Delta_\text{v}^-$. This means that  the deformation factors of the transmission and reflection amplitudes are  the same except for 
a sign $(-1)^M$:
\begin{equation*}
\frac{t_{\mathcal D}(k;\bm{\lambda})}{t(k;\bm{\lambda})}=(-1)^M\frac{r_{\mathcal D}(k;\bm{\lambda})}{r(k;\bm{\lambda})}=
\prod_{j=1}^M\frac{k-i\Delta_{d_j}^-}{k+i\Delta_{d_j}^-}.
\end{equation*}

\subsection{Hyperbolic symmetric top $\II$}
\label{sec:hst}

\paragraph{Original system}

The system has finitely many discrete eigenstates
$0\le n\le n_\text{max}(\bm{\lambda})=[h]'$ in the specified parameter range:
\begin{align*}
  &\bm{\lambda}=(h,\mu),\quad \bm{\delta}=(-1,0),\quad
  -\infty<x<\infty,\quad h,\mu>0,\\
  &U(x;\bm{\lambda})=\frac{-h(h+1)+\mu^2+\mu(2h+1)\sinh x}{\cosh^2x},\\
  &\mathcal{E}_n(\bm{\lambda})=-(h-n)^2,\quad\eta(x)=\sinh x,\\
  &\phi_n(x;\bm{\lambda})
  =i^{-n}e^{-\mu\tan^{-1}\sinh x}(\cosh x)^{-h}\,P_n^{(\alpha,\beta)}(i\eta), \quad W_+=-W_-=h,\\
&\alpha=-h-\tfrac12-i\mu,\ \ \beta=-h-\tfrac12+i\mu.
\end{align*}
The scattering amplitudes of this system are given in \cite{KS}:
\begin{align}
t(k;\bm{\lambda})&=\frac{\Gamma(-h-ik)\Gamma(1+h-ik)\Gamma(\frac12+i\mu-ik)\Gamma(\frac12-i\mu-ik)}
{\Gamma(-ik)\Gamma(1-ik)[\Gamma(\frac12-ik)]^2},
\label{hstt}\\
r(k;\bm{\lambda}) &=  t(k;\bm{\lambda})\left[\cos(\pi h)\sinh(\pi \mu)\,{\rm sech}(\pi k)+ i\sin(\pi h)\cosh(\pi \mu)\,{\rm cosech}(\pi k)\right].
\label{hstr}
\end{align}
The poles on the positive imaginary $k$-axis coming from the first Gamma function factor in the numerator 
of $t(k;\bm{\lambda})$ \eqref{solt}, $-h-i{k}=-n$, $\Rightarrow k=i(h-n)$,  $n=0,1,\ldots, [h]'$ 
provide the eigenspectrum as above. 
In the $\mu\to0$ limit, the potential, eigenvalues, eigenfunctions,
 scattering amplitudes and the polynomial type 
seed solutions  give the corresponding quantities of  soliton potential. 
In this connection, see \S3.9 of \cite{os29}.
 Note that $|t(k;\bm{\lambda})|^2+|r(k;\bm{\lambda})|^2=1$ for $k\in\mathbb{R}_{\ge0}$.
 The potential and the scattering amplitudes \eqref{hstt}--\eqref{hstr} are invariant under the
discrete transformation $h\to -(h+1)$ and $\mu\to -\mu$, but the eigenvalues and the eigenfunctions are not.

It is interesting to note that $t(k;\bm{\lambda})$ and $r(k;\bm{\lambda})$ 
have poles at $k=\pm\mu -(n+\frac12)i$, 
$n\in\mathbb{Z}_{\ge0}$.
In the case of $k=\mu -(n+\frac12)i$ with positive real part, these poles correspond to the so-called quasinormal modes,
whose  real part representing
the actual frequency of oscillation and the imaginary part
representing the damping (in time) \cite{QNM}. 
Such mode have attracted  much interest 
in recently years in the studies of neutron stars and black holes,
as they are produced mainly during the
formation phase of the compact stellar objects, which  may be strong enough to be detected by several large
gravitational wave detectors under construction.
Our results in this subsection indicates that infinitely many quantal systems admitting quasinormal modes 
are provided by the 
multi-indexed extensions of the hyperbolic symmetric top systems.

\paragraph{Polynomial type seed solutions}

The discrete symmetry $h\to -h-1$, $\mu\to-\mu$ generates the pseudo virtual state wavefunctions,
which always lie below the groundstate:
\begin{align}
  &\tilde{\phi}_{\text{v}}(x;\bm{\lambda})
  =e^{\mu\tan^{-1}\sinh x}(\cosh x)^{h+1}
  P_{\text{v}}^{(-{\alpha},-{\beta})}(i\eta),
 \quad
  \tilde{\mathcal{E}}_{\text{v}}(\bm{\lambda})
  =\mathcal{E}_{-\text{v}-1}(\bm{\lambda})<\mathcal{E}_0(\bm{\lambda}),  \n\\
& \text{pseudo virtual}:\quad \Delta_{\text{v}}^+=h+1+\text{v}>0,\quad
\Delta_{\text{v}-}=-\Delta_{\text{v}}^+<0
 \quad(\text{v}\in\mathbb{Z}_{\ge0}).
\end{align}
The overshoot eigenfunctions  for $\text{v}>2h$ provide the 
`pseudo' virtual state wavefunctions:
\begin{align}
  &\tilde{\phi}^{\text{os}}_{\text{v}}(x;\bm{\lambda})
  =\phi_{\text{v}}(x;\bm{\lambda}),\quad
 \tilde{\mathcal{E}}^{\text{os}}_{\text{v}}(\bm{\lambda})
  =\text{v}(2h-\text{v})  
  \n\\
& \text{`pseudo' virtual}:
 \Delta_{\text{v}}^+=\text{v}-h>0,\quad
 \Delta_{\text{v}}^-=-\Delta_{\text{v}}^+<0
 \quad(\text{v}>2h).
\end{align}

\paragraph{Deformed scatterings}
The general properties of the deformed scatterings of soliton potential are shared by hyperbolic symmetric
top, too.

\section{Original and Extended Scatterings: \\
Half line scattering}
\label{sec:half}
This Group (B) consists of three potentials: Morse (M), Eckart (E) and hyperbolic P\"oschl-Teller (hPT).
Although the domain of Morse potential extends from $-\infty$ to $+\infty$, 
the exponentially growing potential
forbids the plane wave to reach $-\infty$ and reflection only occurs.
The reflection amplitudes are determined by the right ($x\to+\infty$) asymptotic properties of the 
solutions which are vanishing at the left boundary.
The discrete symmetry transformations change the  left boundary ($x=0$ or $x=-\infty$)
behaviours.
Thus, {\em the reflection amplitudes are not invariant\/} under the discrete symmetry transformation of the potential.
This is in good  contrast to the full line case in which the transmission amplitudes and reflection amplitudes are
invariant under the discrete symmetry transformations of the potential
In the full line scattering cases, the scattering amplitudes are determined based on 
the two dimensional solution space, which is invariant under the discrete symmetry transformations.
Let us emphasise again that all the original scattering amplitudes presented in this section satisfy the
shape invariance constraints.

\subsection{Morse potential}
\label{sec:M}

\paragraph{Original system}
 The potential and the functions are changed from those of \cite{os28} and \cite{os29}, by $x \to -x$.
This system is defined on the whole line, but there are no transmitted waves to the right because of the
infinitely rising exponential components.  So only the reflection amplitude is defined \cite{KS}.

The system has finitely many discrete eigenstates
$0\le n\le n_\text{max}(\bm{\lambda})=[h]'$ in the specified parameter range:
\begin{align*}
  &\bm{\lambda}=(h,\mu),\quad \bm{\delta}=(-1,0),\quad
  -\infty<x<\infty,\quad h>1/2,\ \mu>0,\\
 & U(x;\bm{\lambda})=\mu^2e^{-2x}-\mu(2h+1)e^{-x},\\
  &\mathcal{E}_n(\bm{\lambda})=-(h-n)^2,\quad\eta(x)=e^{x},\\
  &\phi_n(x;\bm{\lambda})
  =(2\mu)^{-n}e^{-(h-n)x-\mu e^{-x}}\,L_n^{(2h-2n)}(2\mu\eta^{-1}),\quad W_+=h.
 \end{align*}
  The reflection amplitude is given by \cite{KS}
 \begin{align}
 r(k;\bm{\lambda})=(2\mu)^{-2ik}\,\frac{\Gamma(2ik)\Gamma(-h-ik)}{\Gamma(-2ik)\Gamma(-h+ik)}.
 \label{Mr}
 \end{align}
 The poles on the positive imaginary $k$-axis coming from the second Gamma function factor in the numerator 
of $r(k;\bm{\lambda})$ \eqref{Mr}, $-h-i{k}=-n$, $\Rightarrow k=i(h-n)$,  $n=0,1,\ldots, [h]'$ provide the eigenspectrum 
as above. 
 Note that $|r(k;\bm{\lambda})|^2=1$ for $k\in\mathbb{R}_{\ge0}$.
 The potential is invariant under the discrete symmetry $h\to -h-1$, $\mu\to-\mu$ transformation, but the
 reflection amplitude \eqref{Mr} and the eigenvalues with eigenfunctions  are not invariant.
 
 \paragraph{Polynomial type seed solutions}
 
 The discrete symmetry $h\to -h-1$, $\mu\to-\mu$ generates the pseudo virtual state wavefunctions,
 which always lie below the groundstate:
\begin{align}
  &\tilde{\phi}_{\text{v}}(x;\bm{\lambda})
  =e^{(h+1+\text{v})x+\mu e^{-x}}
  L_{\text{v}}^{(-2h-2-2\text{v})}(-2\mu\eta^{-1}),
 \quad  \tilde{\mathcal{E}}_{\text{v}}(\bm{\lambda})
  =\mathcal{E}_{-\text{v}-1}(\bm{\lambda})<\mathcal{E}_0(\bm{\lambda}),\n\\
 &\text{pseudo virtual}:
 \Delta_{\text{v}}^+=(h+1+\text{v})>0
  \quad(\text{v}\in\mathbb{Z}_{\ge0}).
\end{align}
The overshoot eigenfunctions provide type $\I$ virtual state wavefunctions
for $\text{v}>2h$ \cite{quesne5}:
\begin{align}
 & \tilde{\phi}^{\text{os}}_{\text{v}}(x;\bm{\lambda})
  =\phi_{\text{v}}(x;\bm{\lambda}),\quad
  \tilde{\mathcal{E}}^{\text{os}}_{\text{v}}(\bm{\lambda})
  =-(\text{v}-h)^2\n\\
 & \text{type I virtual}:
  \Delta_{\text{v}}^+=\text{v-h}>0
 \quad(\text{v}>2h).
\end{align}
Since this type $\I$ virtual state wavefunction satisfies the boundary
conditions
\begin{equation*}
  \partial_x^s\tilde{\phi}^{\text{os}}_{\text{v}}(x;\bm{\lambda})
  \bigm|_{x=-\infty}=0\quad(s=0,1,\ldots),
\end{equation*}
a multiple virtual state $\tilde{\phi}^{\text{os}}_{\text{v}}$ deletion gives
a non-singular Hamiltonian $\mathcal{H}^{[M]}$.

\paragraph{Deformed scatterings}

As shown above, the overshoot type I virtual state wavefunction has $\Delta_{\text{v}}^+>0$
and it adds a new pole on the positive imaginary $k$-axis, although the corresponding 
Darboux transformation is iso-spectral.
In fact, this new pole at $k=i(\text{v}-h)$ is cancelled by the zero coming from the second Gamma function
in the denominator:
\begin{equation*}
\frac1{\Gamma(-h+ik)}\propto (-h+\text{v}+ik),\quad k\to i(\text{v}-h),\quad \text{v}\in\mathbb{Z}_{\ge0}, \quad
\text{v}>2h.
\end{equation*}
As the other cases,  a pseudo  virtual  state wavefunction will add a new discrete eigenstate at its energy.

\subsection{Eckart potential}
\label{sec:eck}

\paragraph{Original system}
This potential is also called the Kepler problem in hyperbolic space \cite{os28}.
It has finitely many discrete eigenstates
$0\le n\le n_\text{max}(\bm{\lambda})=[\sqrt{\mu}-g]'$ in the specified
parameter range:
\begin{align}
  &\bm{\lambda}=(g,\mu),\quad \bm{\delta}=(1,0),\quad
  0<x<\infty,\quad \sqrt{\mu}>g>\frac32,\n\\
 & U(x;\bm{\lambda})=\frac{g(g-1)}{\sinh^2x}
  -2\mu \left(\coth x -1\right),\n\\
  &\mathcal{E}_n(\bm{\lambda})=-\left[g+n-\frac{\mu}{g+n}\right]^2,\quad\eta(x)=\coth x,\n\\
  &\phi_n(x;\bm{\lambda})
  =e^{-\frac{\mu}{g+n}x}(\sinh x)^{g+n}\,P_n^{(\alpha_n,\beta_n)}(\eta),\quad W_+=\alpha_0,\n\\
  &\alpha_n=-g-n+\frac{\mu}{g+n},\ \ \beta_n=-g-n-\frac{\mu}{g+n}.
  \label{Eckart}
  \end{align}
The reflection amplitude is 
\begin{align}
r(k;\bm{\lambda}) &=  \frac{\Gamma(ik)\Gamma(g-i\frac{k}{2}+
i\frac{k^\prime}{2})\Gamma(g-i\frac{k}{2}-i\frac{k^\prime}{2})}
{\Gamma(-ik)\Gamma(g+i\frac{k}{2}+i\frac{k^\prime}{2})
\Gamma(g+i\frac{k}{2}-i\frac{k^\prime}{2})}.
\label{Ecr}
\end{align}
This functional form is slightly different from that given in \cite{KS}; the present form satisfies
 the shape invariance constraint \eqref{halfshape}. It is a meromorphic function of $k$ and  $k'\eqdef\sqrt{k^2-4\mu}$. 
 
Again we choose the Riemann sheet such that $k'\to k$ in the limit of $\mu\to0$.
The poles on the positive imaginary $k$-axis coming from the second Gamma function factor in the numerator of 
$r(k;\bm{\lambda})$ \eqref{Ecr}, $g-i\frac{k}2+i\frac{k'}2=-n$, $\Rightarrow k=i\alpha_n$, $\alpha_n>0$, 
$n=0,1,\ldots, [\sqrt{\mu}-g]'$ provide the eigenspectrum as above.
It is easy to see  that $|r(k;\bm{\lambda})|^2=1$ for $k\in\mathbb{R}_{\ge0}$. 
The potential is invariant under the discrete symmetry $g\to 1-g$, $\mu\to-\mu$ transformation, but the
 reflection amplitude \eqref{Ecr} and the eigenvalues with eigenfunctions  are not invariant.

\paragraph{Polynomial type seed solutions}
  
 The discrete symmetry $g\to 1-g$, $\mu\to\mu$ generates the pseudo virtual and virtual state wavefunctions
  \cite{quesne5}:
\begin{align}
  &\tilde{\phi}_{\text{v}}(x;\bm{\lambda})
  =e^{\frac{\mu}{g-\text{v}-1}x}(\sinh x)^{-g+\text{v}+1}
  P_{\text{v}}^{(\bar{\alpha}_\text{v},\bar{\beta}_\text{v})}(\eta),\n\\
& \Delta_{\text{v}}^+=-\bar{\alpha}_\text{v} \qquad
\left\{
 \begin{array}{lll}
  \text{pseudo virtual}:&
  {\displaystyle 0<\text{v}<g-1},&\Delta_{\text{v}}^+>0\\
    \text{type $\II$ virtual}:&
  {\displaystyle g-1<\text{v}<2g-1},&\Delta_{\text{v}}^+<0\\
   \text{pseudo virtual}:&
  {\displaystyle  \text{v}>\frac{\mu}{g}+g-1},&\Delta_{\text{v}}^+>0
  \end{array}\right.,\\
  &\bar{\alpha}_\text{v}=g-\text{v}-1-\frac{\mu}{g-\text{v}-1},\ \ \bar{\beta}_\text{v}=g-\text{v}-1+\frac{\mu}{g-\text{v}-1},\\
  &\tilde{\mathcal{E}}_{\text{v}}(\bm{\lambda})
  =\mathcal{E}_{-\text{v}-1}(\bm{\lambda})
  \quad\bigl(0\le\text{v}<g-1,\ \text{v}>\frac{\mu}{g}+g-1\bigr).
\end{align}

The overshoot eigenfunctions  provide type $\I$ virtual
state wavefunctions for $\text{v}>\frac{\mu}{g}-g$  \cite{quesne4}:
\begin{align}
  &\tilde{\phi}^{\text{os}}_{\text{v}}(x;\bm{\lambda})
  =\phi_{\text{v}}(x;\bm{\lambda}),\quad
  \tilde{\mathcal{E}}^{\text{os}}_{\text{v}}(\bm{\lambda})
  =\mathcal{E}_\text{v}(\bm{\lambda}),\n\\
 & \text{type I virtual}: \Delta_{\text{v}}^+=-\alpha_\text{v}>0
    \quad\bigl(\text{v}>\text{max}({\mu}/{g}-g, 2g-1)\bigr).
\end{align}
Since this type $\I$ virtual state wavefunction satisfies the boundary
conditions
\begin{equation*}
  \partial_x^s\tilde{\phi}^{\text{os}}_{\text{v}}(x;\bm{\lambda})
  \bigm|_{x=0}=0\quad(s=0,1,\ldots),
\end{equation*}
a multiple virtual state $\tilde{\phi}^{\text{os}}_{\text{v}}$ deletion gives
a non-singular Hamiltonian $\mathcal{H}^{[M]}$.

\paragraph{Deformed scatterings}
The overshoot type I virtual state wavefunction has $\Delta_{\text{v}}^+>0$
and it adds a new pole on the positive imaginary $k$-axis, although the corresponding 
Darboux transformation is iso-spectral.
In fact, this new pole at $k=-i\alpha_\text{v}$ is cancelled by the zero coming from the second Gamma function
in the denominator:
\begin{equation*}
 \frac1{\Gamma(g+i\frac{k}{2}+i\frac{k^\prime}{2})}\propto [2(g+\text{v})+ik+ik']/2,\quad k\to -i\alpha_\text{v},\quad
 k'\to-i\beta_\text{v}.
\end{equation*}
The type II virtual state wavefunction, on the other hand, will add a pole on the negative imaginary $k$-axis,
which does not correspond to a bound state.
The  pseudo  virtual  state wavefunction will add a new discrete eigenstate at its energy.

\subsection{Hyperbolic P\"{o}schl-Teller potential}
\label{sec:hDPT}

\paragraph{Original system}

This potential has finitely many discrete eigenstates
$0\le n\le n_\text{max}(\bm{\lambda})=[\frac{h-g}{2}]'$ in the specified
parameter range:
\begin{align*}
  &\bm{\lambda}=(g,h),\quad \bm{\delta}=(1,-1),\quad
  0<x<\infty,\quad h>g>\frac12,\\
  &
  U(x;\bm{\lambda})=\frac{g(g-1)}{\sinh^2x}
  -\frac{h(h+1)}{\cosh^2 x},\\
  &\mathcal{E}_n(\bm{\lambda})=-(h-g-2n)^2,\quad\eta(x)=\cosh 2x,\\
  &\phi_n(x;\bm{\lambda})
  =(\sinh x)^g(\cosh x)^{-h}\,P_n^{(g-\frac12,-h-\frac12)}(\eta),\quad W_+=h-g.
\end{align*}
The reflection amplitude is 
\begin{equation}
r(k;\bm{\lambda}) = 2^{-2ik} \frac{\Gamma(ik)\Gamma\bigl((-h + g -ik)/2)\Gamma\bigl((1+h+g-ik)/2\bigr)}
{\Gamma(-ik)\Gamma\bigl((-h + g +ik)/2\bigr)\Gamma\bigl((1+h+g+ik)/2\bigr)},
\label{hPTr}
\end{equation}
which is different from that given in \cite{KS} by the factor $2^{-2ik}$.
The poles on the positive imaginary $k$-axis coming from the second Gamma function factor in the numerator 
of $r(k;\bm{\lambda})$ \eqref{Mr}, $-h+g-i{k}=-2n$, $\Rightarrow k=i(h-g-2n)$,  $n=0,1,\ldots, [\frac{h-g}{2}]'$ 
provide the eigenspectrum  as above.  Note that $|r(k;\bm{\lambda})|^2=1$ for $k\in\mathbb{R}_{\ge0}$. 
It is interesting to note that the potential is invariant under the discrete symmetry transformation,
$g\to 1-g$, or  $h\to -h-1$,  or $g\to 1-g$ and $h\to -h-1$,
 but the
 reflection amplitude \eqref{hPTr} is invariant only under the transformation $h\to -h-1$, 
 which does not change the left boundary condition. The eigenvalues and eigenfunctions  are not invariant.

 \paragraph{Polynomial type seed solutions}

Two types of  virtual and pseudo virtual state wavefunctions are generated by the discrete symmetry transformations:
\begin{align} 
&\text{type I virtual}: \quad h\to -h-1 \quad  \tilde{\mathcal E}_\text{v}(\bm{\lambda})=-(h+1+g+2\text{v})^2<\mathcal{E}_0(\bm{\lambda}),\n\\
& \hspace{32mm}
\phi_\text{v}(x;\bm{\lambda})
  =(\sinh x)^g(\cosh x)^{h+1}\,P_\text{v}^{(g-\frac12,h-\frac12)}(\eta),\n\\
 &  \hspace{32mm} \Delta_{\text{v}}^+=g+h+1+2\text{v}>0\quad (\text{v}\in\mathbb{Z}_{\ge0}),
  \label{hDPTI}\\
&\text{type II virtual}: \quad g\to 1-g \quad\  \tilde{\mathcal E}_\text{v}(\bm{\lambda})=-(h+g-1-2\text{v})^2, \n\\
&
\tilde{\phi}_\text{v}(x;\bm{\lambda})
  =(\sinh x)^{1-g}(\cosh x)^{-h}\,P_\text{v}^{(\frac12-g,-h-\frac12)}(\eta),
  \quad \tilde{\mathcal E}_\text{v}(\bm{\lambda})=-(h+g-1-2\text{v})^2,
  \n\\[2pt]
 &  \Delta_{\text{v}}^+=1-g-h+2\text{v}\qquad
\left \{
\begin{array}{lll}
  \text{type II virtual}:&  \Delta_{\text{v}}^+<0 & 0<\text{v}<g-\frac12  \\[4pt]
 \text{pseudo virtual}: &  \Delta_{\text{v}}^+>0 &  \text{v}>h-\frac12   
\end{array}
\right..
\end{align}

The type II virtual states are below the groundstate energy for $0<\text{v}<g-1/2$.
Pseudo virtual state wavefunctions are generated by $g\to1-g$, $h\to-h-1$:
\begin{align}
  &\tilde{\phi}_{\text{v}}(x;\bm{\lambda})
  =(\sinh x)^{1-g}(\cosh x)^{h+1}
  P_{\text{v}}^{(\frac12-g,h-\frac12)}(\eta)
 \quad
\tilde{\mathcal{E}}_{\text{v}}(\bm{\lambda})
  =\mathcal{E}_{-\text{v}-1}(\bm{\lambda}),\n\\
&\text{pseudo virtual}: \Delta_{\text{v}}^+=h-g+2+2\text{v}>0
 \quad(\text{v}\in\mathbb{Z}_{\ge0}).
\end{align}
The overshoot eigenfunctions  provide type $\I$ virtual state
wavefunctions for $\text{v}>h-g$:
\begin{align}
  &\tilde{\phi}^{\text{os}}_{\text{v}}(x;\bm{\lambda})
  =\phi_{\text{v}}(x;\bm{\lambda}),\quad
  \tilde{\mathcal{E}}^{\text{os}}_{\text{v}}(\bm{\lambda})
  =\mathcal{E}_\text{v}(\bm{\lambda})=-(h-g-2\text{v})^2, \n\\
& \text{type I virtual}: \Delta_{\text{v}}^+=g-h+2\text{v}>0   \quad(h-g<\text{v}<h+\frac12).
 \label{hDPTI2}
\end{align}

\paragraph{Deformed scatterings}

The  type I virtual state wavefunction \eqref{hDPTI} has $\Delta_{\text{v}}^+>0$
and it adds a new pole on the positive imaginary $k$-axis, although the corresponding 
Darboux transformation is iso-spectral.
In fact, this new pole at $k=i(g+h+1+2\text{v})$ is cancelled by the zero coming from the third Gamma
function in the denominator:
\begin{equation*}
\frac1{\Gamma\bigl((1+h+g+ik)/2\bigr)}\propto (g+h+1+2 \text{v}+ik)/2,\quad k\to i(g+h+1+2\text{v}),\quad \text{v}\in\mathbb{Z}_{\ge0}.
\end{equation*}
Another pole created by the overshoot type I virtual state wavefunction \eqref{hDPTI2}  at 
$k=i(g-h+2\text{v})$ is cancelled by the zero coming from the second Gamma
function in the denominator:
\begin{equation*}
\frac1{\Gamma\bigl((-h+g+ik)/2\bigr)}\propto (g-h+2 \text{v}+ik)/2,\quad k\to i(g-h+2\text{v}),\quad 
(h-g<\text{v}<h+\frac12).
\end{equation*}
The type II virtual state wavefunction, on the other hand, will add a pole on the negative imaginary $k$-axis,
which does not correspond to a bound state.

\section{Coulomb potential plus the centrifugal barrier (C)}
\label{sec:coul}

Coulomb potential plus the centrifugal barrier is defined in the interval $0<x<+\infty$, 
but it needs to be separately discussed from the others because of its long range character.

\paragraph{Original system}

Unlike the other systems presented in this paper, the Coulomb system has  
infinitely many discrete eigenstates in the specified parameter range:
\begin{align*}
  &\bm{\lambda}=g,\quad \bm{\delta}=1,\quad
  0<x<\infty,\quad g>\frac32,\\
  &
  U(x;\bm{\lambda})=\frac{g(g-1)}{x^2}-\frac{2}{x},\\
  &\mathcal{E}_n(\bm{\lambda})=-\frac{1}{(g+n)^2},\quad n\in\mathbb{Z}_{\ge0},\quad
  \eta(x)=x^{-1},\\
  &\phi_n(x;\bm{\lambda})
  =e^{-\frac{x}{g+n}}x^{g}\,L_n^{(2g-1)}\bigl(\tfrac{2}{g+n}\eta^{-1}\bigr), \quad W_+=\frac1g.
  \end{align*}
The scattering amplitude is given by:
\begin{equation}
r(k;\bm{\lambda})=e^{-i\pi g}\frac{\Gamma(g-\frac{i}{k})}{\Gamma(g+\frac{i}{k})}
=e^{-i\pi g}\frac{\Gamma(g+i\gamma)}{\Gamma(g-i\gamma)},
\label{coulr}
\end{equation}
in which the constant $\gamma=-1/k$ enters in the asymptotic form of the wavefunction \eqref{coulasym}.
The phase factor $e^{-i\pi g}$ is necessary to satisfy the shape invariance constraint \eqref{halfshape}.
 It is important to note that this factor survives in the 
calculation of 1-dimensional Coulomb scattering, but not in the standard 3-dimensional
Coulomb scattering.
The reflection amplitude has infinitely many poles $k=i/(g+n)$, $n\in\mathbb{Z}_{\ge0}$, 
on the positive imaginary $k$-axis.
In \cite{KS}, the reported scattering amplitude 
$r(k)$ has extra factors $\Gamma(1+i/k)/\Gamma(1-i/k)$ instead of $e^{-i\pi g}$. Because of the extra gamma function factors, their reflection amplitude
becomes trivial $r(k)=1$ at $g=1$, at which the potential is {\em not free}. 
It is easy to see  that $|r(k;\bm{\lambda})|^2=1$ for $k\in\mathbb{R}_{>0}$. 
The potential is invariant under the discrete symmetry $g\to 1-g$ transformation, but the
 reflection amplitude \eqref{coulr} and the eigenvalues and eigenfunctions  are not invariant.

 \paragraph{Polynomial type seed solutions}
 
The discrete symmetry $g\to 1-g$ generates the pseudo virtual  state wavefunctions 
\cite{grandati}:
\begin{align}
  &\tilde{\phi}_{\text{v}}(x;\bm{\lambda})
  =e^{\frac{x}{g-\text{v}-1}}x^{1-g}
  L_{\text{v}}^{(1-2g)}\bigl(\tfrac{2}{1-g+\text{v}}\eta^{-1}\bigr) 
  \quad \tilde{\mathcal{E}}_{\text{v}}(\bm{\lambda})
    =\mathcal{E}_{-\text{v}-1}(\bm{\lambda})<\mathcal{E}_0(\bm{\lambda}),  \n\\
&\qquad
x\to+\infty \qquad \tilde{\phi}_{\text{v}}(x;\bm{\lambda})\approx e^{\frac{x}{g-\text{v}-1}}x^{1-g+\text{v}},
 \label{coulasym2}\\
 &   \Delta_{\text{v}}^+=\frac{1}{g-\text{v}-1}\qquad
\left\{
\begin{array}{lll}
 \text{pseudo virtual}: &  \Delta_\text{v}^+>0 &0\le\text{v}<g-1   \\[2pt]
  \text{type II virtual}: &    \Delta_\text{v}^+<0 &g-1<\text{v}<2g-1 
\end{array}
\right..
\end{align}
In contrast with the other potentials, the asymptotic form of the polynomial type seed solutions
\eqref{coulasym2} has an extra power function  of $x$  multiplied to the asymptotic exponents \eqref{asymcouls}.

\paragraph{Deformed scatterings}

A type II virtual state wavefunction would give a pole on the negative imaginary $k$-axis and a zero 
with the opposite sign. As the others,  the pseudo virtual state wavefunction produces a pole on the positive imaginary
$k$-axis.

%
%
\section{Summary}
\label{sec:sum}

The method of multi-indexed extensions has been successfully applied to confining and semi-confining solvable 
potentials, revealing the interesting properties of the deformed discrete eigenfunctions; {\em e.g.\/}
the multi-indexed orthogonal polynomials and the exceptional orthogonal polynomials.
In this paper, scattering problem aspects of the same deformed systems are investigated.
The results are surprisingly rich, simple and beautiful.
The asymptotic behaviours of each polynomial type seed solution are characterised by two
 (full line) or one (half line) {\em asymptotic exponents\/} $\Delta_\text{v}^\pm$ or $\Delta_\text{v}^+$, 
 \eqref{asymfulls}--\eqref{asymcouls}, 
 the explicit values of these
 exponents are given for each solvable potential in sections four, five and six.
 The transmission amplitude $t(k)$ (full) and the reflection 
 amplitude $r(k)$ (full \& half) obtain a new pole and a new zero on the imaginary $k$-axis determined 
 by the asymptotic exponents, \eqref{trDfull}--\eqref{trDRM}, \eqref{rBC}.
 The combinations of various polynomial type seed solutions determine the regularity or singularity of the
 deformed potential \eqref{vham}, which is simply given by the second logarithmic derivatives of 
 the Wronskian of 
 all the seed solutions. Various criteria for the non-singularity of the deformed potentials are provided.
 Two papers discussing similar targets but with limited contents appeared recently \cite{yadav}.
 Extension to QNM is also discussed.
 
 It would be interesting to pursue the same problem from a different angle;
 deformed scattering problem of the same solvable potentials, using the same or generalised 
 polynomial type seed solutions but by applying multiple Abraham-Moses transformations \cite{A-M,os31}.

\section*{Acknowledgements}
R.\,S. thanks Pei-Ming Ho for the hospitality at
National Center for Theoretical Sciences (North), National Taiwan University.
C.L.H. is supported in part by the
National Science Council (NSC) of the Republic of China under
Grants NSC-99-2112-M-032-002-MY3 and NSC-102-2112-M-032-003-MY3.  J.C.L. is support in part by NSC-100-2112-M-009-002-MY3, 
the 50 billions(NTD) project of the Ministry of 
Education (Taiwan) and S.T. Yau center of NCTU, Taiwan. 
R.\,S. is supported   in part by Grant-in-Aid for Scientific Research
from the Ministry of Education, Culture, Sports, Science and Technology
(MEXT) No.22540186.
We also acknowledge the support by the National Center for Theoretical Sciences (NCTS) of R.O.C.


 \end{document}